\documentclass[11pt]{article}

\usepackage{graphicx}
\usepackage{amssymb,amsmath}
\usepackage{color}
\usepackage[compact,small]{titlesec}
\titleformat{\section}[block]
  {\large\bfseries\filright}
  {\thesection}{6pt}{}
\titleformat{\subsection}[runin]
  {\normalfont\bfseries}
  {\thesubsection}{6pt}{}
\titleformat{\subsubsection}[runin]
  {\normalfont\bfseries}
  {\thesubsubsection}{6pt}{}

\newenvironment{itemize*}{
\begin{itemize}
\setlength{\parskip}{0em}
\setlength{\topskip}{0em}
}
{\end{itemize}}

\newenvironment{enumerate*}{
\begin{enumerate}
\setlength{\parskip}{0em}
\setlength{\topparskip}{0em}
}
{\end{enumerate}}

\newcommand{\beq}{\begin{equation}}
\newcommand{\eeq}{\end{equation}}
\newcommand{\beqa}{\begin{eqnarray}}
\newcommand{\eeqa}{\end{eqnarray}}
\newcommand{\beqas}{\begin{eqnarray*}}
\newcommand{\eeqas}{\end{eqnarray*}}

\newcommand{\bit}{\begin{itemize}}
\newcommand{\eit}{\end{itemize}}
\newcommand{\bits}{\begin{itemize*}}
\newcommand{\eits}{\end{itemize*}}
\newcommand{\benum}{\begin{enumerate}}
\newcommand{\eenum}{\end{enumerate}}
\newcommand{\benums}{\begin{enumerate*}}
\newcommand{\eenums}{\end{enumerate*}}
\newcommand{\comment}[1]{}

\newcommand{\todo}[1]{}%{{\textcolor{red}{\em TODO: #1}}}
\newcommand{\mmp}[1]{}%{\textcolor{blue}{\em MMP: #1}}

\newcommand{\mf}[1]{}%{\textcolor{green}{MF: #1}}

\newcommand{\clust}{{\cal C}}

\newcommand{\T}{{\mathcal T}}
\newcommand{\scpalg}{{\sc SampleConnectedPartition}}

\title{How to sample connected $K$-partitions of a graph}
\author{Marina Meila\\{\tt mmp@stat.washington.edu}}

\begin{document}
\maketitle

\begin{abstract}
A connected undirected graph $G=(V,E)$ is given. This paper presents an algorithm that samples (non-uniformly) a $K$ partition $U_1,\ldots U_K$ of the graph nodes $V$, such that the subgraph induced by each $U_k$, with $k=1:K$, is connected. Moreover, the probability induced by the algorithm over the set $\clust_K$ of all such partitions is obtained in closed form.
\end{abstract}

\section{Problem and notation}
A connected undirected graph $G=(V,E)$ with $|V|=n$ is given. A {\em connected $K$-partition} of $G$ denotes  a  partition of $V$ into $K$ {\em clusters} $U_1,\ldots U_K$, such that the subgraph of $G$ induced by each $U_k$, with $k=1:K$, is connected. Here $K$ is considered fixed and may be omitted for brevity. 

A connected partition is denoted by $C$, and the set of all connected $K$ partitions of $G$ is denoted by $\clust_K$. Counting $|\clust_K|$ is known to be hard in general \cite{vince:17}. 

Denote by $T$ a spanning tree of $G$, and by $\T$ the set of all spanning trees of $G$. The spanning trees of a simple undirected graph can be counted by Tutte's {\em Matrix Tree Theorem} \cite{West:01}. This theorem extends to multigraphs with no self loops. Let $t(G)=|\T|$, and $t(S)$ the number of spanning trees in the subgraph of $G$ induced by $S\subset V$. The Matrix Tree Theorem states that $t(G)=\det(L(G)^*)$ where $L(G)=D(G)-A(G)$ the diagonal degree matrix minus the adjacency matrix of $G$ (i.e. the unnormalized Laplacian of graph $G$), and $L^*$ is a minor of matrix $L$, i.e $L$ with the $i$-th row and column removed, for some arbitrary $i$. Note that $t(G)$ is 0 if $G$ is not connected and that $\det L=0$ always, as the rows of $L$ sum to 0.

\section{An algorithm for sampling from $\clust_K$}

The following algorithm samples connected $K$-partions, {\em non-uniformly}.

\benum
\item[] Algorithm \scpalg$(K,G)$
\item Sample a spanning tree $T\in\T$ uniformly at random. 
\item Remove $K-1$ edges from $T$ uniformly  at random without replacement. 
\item[]Return the connected components $U_{1:K}$ of $T$ obtained in Step 2.
\eenum

Proof sketch: it is obvious that each $U_k$ is connected. 
Step 1 can be performed for example by assigning the edges random weights and computing the minimum spanning tree with these weights. 

We say that a spanning tree $T\in \T$ is {\em compatible} with a partition $C\in \clust_K$ iff $C$ can be obtained from $T$ by removing $K-1$ edges.

\section{Analysis. Probability induced by \scpalg~on $\clust_K$}

The question now is: what is the probability of obtaining a given partition $U_{1:K}$ by the \scpalg~algorithm? 

We first explain the idea for $K=2$; in this case we remove a single
edge from $T$. Let $S\subset V$ ($S$ represents $U_1$ or
$U_2$). Denote by $\partial S$ the edges between $S$ and $V\setminus
S$. Any spanning tree $T$ must intersect $\partial S$ (otherwise $T$
would not be connected). If $|T\cap \partial S|>1$, no edge removal
will produce the partition $C=(S,V\setminus S)$. But if $|T\cap \partial S|=1$, then w.p. $1/(n-1)$ the partition is obtained, namely
when the single edge in $T\cap \partial S$ is deleted from $T$.

For a fixed $S$, let the event $\T_S=\{|T\cap \partial S|=1\}\subset \T$. Note that fixing $S$ in this case amounts to fixing the partition $C$. 

Any $T$ in $\T_S$ contains a spanning tree of $S$, a spanning tree of $V\setminus S$, and one edge from $\partial S$. Hence, 
\beq
|\T_S|\;=\;t(S)t(V\setminus S)|\partial S|
\eeq
and 
\beq
P(C)\;=\;\frac{P(\T_S)}{n-1}\;=\;\frac{t(S)t(V\setminus S)|\partial S|}{(n-1)t(G)}
\eeq
Now, let's consider the general case of a $K$ partition $C=(U_1,\ldots U_K)$. Each $T$ that is compatible with $C$ must contain a spanning tree $T_k$ of the subgraph induced by $U_k$, for each $k=1:K$. Furthermore, these trees must be connected by edges between two clusters $U_k,U_{k'}$, ensuring that no loops are formed. In other words, to complete $\cup_{1:k}T_k$ to a spanning tree $T$ of $G$ that is compatible with $C$, we {\em contract} each $U_k$ to a single node; all the edges between $U_k$ and $U_{k'}$ are now between the two nodes representing $U_k$ and $U_{k'}$. Hence, we obtain a {\em multigraph} $M(G,C)$ with $K$ nodes. Any spanning tree of $M(G,C)$ completes $\cup_{1:k}T_k$ to a spanning tree of $G$.

The number of spanning trees in the multigraph $M(G,C)$ is obtained again by the Matrix Tree Theorem, where each edge has a weight equal to its multiplicity.

Once we have a $T$ compatible with $C$, we need to remove the set of $K-1$ edges connecting the clusters $U_{1:K}$, out of $\binom{n-1}{K-1}$ possible edge removals. Hence,
\beq
P(U_{1:K})\;=\;\frac{t(M(G,U_{1:K}))\prod_{k=1}^Kt(U_k)}{\binom{n-1}{K-1}t(G)}.
\eeq
This analysis also shows that \scpalg~samples every connected partition of $G$ with non-zero probability.
\section{An example}

Let the graph $G$ with $n=10$ be defined by the following adjacency matrix $A$.
\begin{center}
\begin{tabular}{|l|r|r|r|r||r|r|r||r|r|r|r|}
%\begin{tabular}{|l|l|l|l|l|l|l|l|l|l|l|}
\hline
&\textbf{ 1}&\textbf{ 2}&\textbf{ 3}&\textbf{ 4}&\textbf{ 5}&\textbf{ 6}&\textbf{ 7}&\textbf{ 8}&\textbf{ 9}&\textbf{10}\\\hline
\textbf{ 1}&0&1&1&1&0&0&0&0&0&0\\\hline
\textbf{ 2}&1&0&1&1&1&0&0&0&0&0\\\hline
\textbf{ 3}&1&1&0&1&0&0&0&0&1&0\\\hline
\textbf{ 4}&1&1&1&0&0&1&0&0&0&0\\\hline\hline
\textbf{ 5}&0&1&0&0&0&1&1&0&0&0\\\hline
\textbf{ 6}&0&0&0&1&1&0&1&0&0&1\\\hline
\textbf{ 7}&0&0&0&0&1&1&0&1&0&0\\\hline\hline
\textbf{ 8}&0&0&0&0&0&0&1&0&1&1\\\hline
\textbf{ 9}&0&0&1&0&0&0&0&1&0&1\\\hline
\textbf{10}&0&0&0&0&0&1&0&1&1&0\\\hline
\end{tabular}
\end{center}
The node degrees are
\begin{center}
\begin{tabular}{|r|r|r|r|r|r|r|r|r|r|r|}

\hline
\textbf{ 1}&\textbf{ 2}&\textbf{ 3}&\textbf{ 4}&\textbf{ 5}&\textbf{ 6}&\textbf{ 7}&\textbf{ 8}&\textbf{ 9}&\textbf{10}\\\hline
3&4&4&4&3&4&3&3&3&3\\\hline
\end{tabular}
\end{center}
and the Laplacian matrix is 
\begin{center}
\begin{tabular}{|r|r|r|r|r||r|r|r||r|r|r|}
\hline
&\textbf{ 1}&\textbf{ 2}&\textbf{ 3}&\textbf{ 4}&\textbf{ 5}&\textbf{ 6}&\textbf{ 7}&\textbf{ 8}&\textbf{ 9}&\textbf{10}\\\hline
\textbf{ 1}&3&-1&-1&-1&0&0&0&0&0&0\\\hline
\textbf{ 2}&-1&4&-1&-1&-1&0&0&0&0&0\\\hline
\textbf{ 3}&-1&-1&4&-1&0&0&0&0&-1&0\\\hline
\textbf{ 4}&-1&-1&-1&4&0&-1&0&0&0&0\\\hline\hline
\textbf{ 5}&0&-1&0&0&3&-1&-1&0&0&0\\\hline
\textbf{ 6}&0&0&0&-1&-1&4&-1&0&0&-1\\\hline
\textbf{ 7}&0&0&0&0&-1&-1&3&-1&0&0\\\hline\hline
\textbf{ 8}&0&0&0&0&0&0&-1&3&-1&-1\\\hline
\textbf{ 9}&0&0&-1&0&0&0&0&-1&3&-1\\\hline
\textbf{10}&0&0&0&0&0&-1&0&-1&-1&3\\\hline
\end{tabular}
\end{center}
Let the $K=3$ clusters be $U_1=\{1,2,3,4\}$, $U_2=\{5,6,7\}$, $U_3=\{8,9,10\}$. 
Then, 
\beq
t(A)\, = \,\det(L_{1:9,1:9})\,=\,4,546
\quad
t(U_1)\,=\,16,
\quad t(U_2)\,=\,t(U_3)\,=\,3
\eeq
and
\beq
M(G)\,=\,\left[\begin{array}{lll} 0& 2& 1\\
                            2& 0& 2\\
                            1& 2& 0\\
                            \end{array}\right]
\quad t(M(G))\,=\,8
\quad \binom{n-1}{K-1}\,=\,36
\eeq
Hence, the probability of the partition $(U_1,U_2,U_3)$ is equal to $\frac{16\times 3\times 3\times 8}{36\times 4546}=0.0070$.

\section*{Acknowledgement} This problem was suggested by a question from Steve K.


\begin{thebibliography}{10}

\bibitem{vince:17}  
A. Vince
\newblock Counting connected sets and connected partitions of a graph
\newblock Australasian Journal of Combinatorics, 67, 2017.

\bibitem{West:01} 
D. M. West
\newblock  An introduction to graph theory
\newblock  Prentice Hall, 2001.
\end{thebibliography}
\end{document}